\documentclass[10pt]{article}
\usepackage[dvips]{graphicx}
\usepackage{amssymb}
\begin{document}
\begin{center}
{ \LARGE {\bf Enlarging Maurer-Cartan form via Kronecker product and construction of Coupled Integrable systems by Nilpotent, Hadamard, Idempotent and K-idempotent matrix.\\\
 } }
\end{center}
\vskip 0pt
\begin{center}
{\it {\large $Arindam \hskip 2 pt Chakraborty$
\footnote {e-mail: arindam.chakraborty@heritageit.edu}  }\\
\it{Department of Physics ,
 Heritage Institute of Technology,\\
                              Calcutta - 700107,
                                    India}  }
\end{center}

\vskip 20pt
\begin{center}
{\bf Abstract}
\end{center}

\par Coupled nonlinear integrable systems are generated from usual zero curvature equation. The relevant Maurer-Cartan forms are constructed by combining suitably chosen matrices (nilpotent, Hadamard, idempotent and k-idempotent) and Lie algebraic elements via Kronecker product. In each case a closure type property among the matrices chosen is found to be playing a key role to produce both the coupling and nonlinearity present in the system of equations obtained. The method is highly flexible and can be used to construct general systems containing 'p' number of equations. It is also shown that these new equations can be written in the Hamiltonian form (with a preassigned symplectic operator) with the trace identity introduced by Tu. Since the Lax operator is known one can obtain the hereditary operators signifying the complete integrability. Various properties of Kronecker product are found to be useful in our construction.

 \par PACS Number(s):  05.45.Pq, 05.45.Ac, 05.45.-a.

\section{Introduction}
Extending the class of integrable equations$^{[1],[2],[3]}$ is one of  the most important areas of research in the field of solitonic equations. Among various avenues those have been explored to obtain continuous and discrete types of flow one can mention the so called symmetric space construction$^{[4]}$, extension of underlying Lie algebra though direct sum$^{[5]}$, construction of new loop algebra$^{[6]}$ to name a few. In this context it may be mentioned that it was B. Fucssteiner$^{[7]}$ who first demonstrated how one can generate coupled integrable systems by starting with the original nonlinear equation and its linearized forms.  Use has also been made of the hereditary operator and Lie symmetry of the equation$^{[8]}$. Extension of such a method was done by Zang et. al. $^{[9]}$. The situation involving infinite dimensional case was treated by Bacocai and Zhuquan $^{[10]}$. Later it was found that by coupling of commutator pairs coupled solitonic system can be manufactured as described by Zhang and Tau$^{[11]}$. It was also observed that by decoupling of a Lie algebra in its various subalgebras one can achieve such a goal$^{[12]}$. Of course one should not forget the method of Whalquist and Estabrook for the derivation of a Lax pair from a given nonlinear system$^{[12],[13],[14],[15]}$. These methods often lead to coupled integrable equations that keep intact the property of integrability. The best way to ascertain the property of integrability is to assure the existence of Lax pairs understood through zero curvature equation that involves derivations in relation to space and time variables. In a recent communication we have shown how a Lax pair can be formed for typical (1+1) dimensional systems with the help of Kronecker product$^{[16]}$ between various projection type operators in one hand  and suitably chosen Lie algebraic elements on the other. As a consequence the basic zero curvature equation leads us to a series of underlying coupled equations at least to within a sense of sufficient condition.
\par In the present communication what has been constructed is the basic 1-forms by combining four distinct types of  matrices (nilpotent, Hadamard, idempotent and k-idempotent matrices) with iso-spectral Lie algebraic elements relating to AKNS or KN type systems$^{[17]}$. This helps us to appreciate our method  in at least two different loop algebraic contexts. One of the interesting observations regarding the use of such matrices is that they are found to hold a kind of closure type relation among themselves: a crucial fact that we have demonstrated in tabular form for the k-idempotent case. This fact along with several other algebraic properties of these matrices greatly influence the nature of coupling present in the nonlinear equations thus obtained. As an example in our first case with nilpotent matrix we find as many nonlinear equations as the index of nilpotency. Entries of the relevant recursion matrices leading to so called hierarchy of nonlinear equation have been enlisted in the appendix for at least two such cases (Hadamard and k-idempotent) discussed.
\par The integrability property of the equations so obtained has been envisaged through the construction of Hamiltonian and by demonstrating equations of motions in the sense of Liouville, each case admitting a distinct symplectic operator. The relevant trace identity$^{[18]}$ leading to the  Hamiltonians and the variety of symplectic operators associated with them demands the use of the properties of Kronecker product and the contribution of various matrices that we have brought into use. In fact the very presence of expressions in the form of Kronecker product lead us to introduce a new form of inner product in the nilpotent case.  It has been appreciated in literature $^{[5], [17], [18]}$ that various types of such inner product provide one of the key points leading to the formulation of relevant Hamiltonians.
\section{Formulation}
\subsection{Enlarging Maurer-Cartan form}
Given n differential i-form $\theta_j,j=1\dots n$ with Lie algebraic coefficients one can construct the following expression
\begin{eqnarray}
\theta=\sum_{j=1}^n M_j\otimes\theta_j
\end{eqnarray}
where $\theta_j=U_jdx+V_jdt$ with $U_j$ and $V_j$ belongs to a chosen Lie algebra.
With this the corresponding Maurer-Cartan form $d\theta-\theta\bigwedge\theta=0$(zero curvature equation) leads to a series of nonlinear equations. In fact this is equivalent to considering a Lax pair $U, V$ with the following demand 
\begin{eqnarray}
U=\sum_{j=1}^n M_j\otimes U_j
\end{eqnarray}
\begin{eqnarray}
V=\sum_{j=1}^n M_j\otimes V_j
\end{eqnarray}
and
\begin{eqnarray}
U_t-V_x+[U, V]=0
\end{eqnarray} 
\subsection{Coupling by nilpotent matrix}
To start with let us consider a set of x-part of Lax matrices $\{U_j\mid j=1\dots p+1\}$ and construct
\begin{eqnarray}
U=\textbf{1} \otimes U_1+\sum_{j=1}^p N^j\otimes U_{j+1}
\end{eqnarray}
where $\{N^j\neq 0\mid j=1\dots p\}$ and $N^{p+1}=0$
Here, $N$ is a nilpotent matrix. Similarly for the time part we take
\begin{eqnarray}
V=\textbf{1}\otimes V_1+\sum_{j=1}^p N^j\otimes V_{j+1}
\end{eqnarray}
We demand the zero curvature condition which leads to
\begin{eqnarray}
U_{1t}-V_{1x}+[U_1,V_1]=0\nonumber\\
U_{n+1t}-V_{n+1x}+\sum_{j=1}^{n+1}[U_j,V_{n-j+2}]=0
\end{eqnarray}
for $\{n=1\dots p\}$.
\par On the other hand a slightly different form can be taken as
\begin{eqnarray}
U=\textbf{1}\otimes U_1+\sum_{j=1}^p(-1)^jN^j\otimes U_{j+1}\nonumber\\
V=\textbf{1}\otimes V_1+\sum_{j=1}^p(-1)^jN^j\otimes V_{j+1}
\end{eqnarray}
Whence the resultant equations are
\begin{eqnarray}
U_{1t}-V_{1x}+[U_1,V_1]=0\nonumber\\
U_{n+1t}-V_{n+1x}+(-1)^n\sum_{j=1}^{n+1}[U_j,V_{n-j+2}]=0
\end{eqnarray}
for $\{n=1\dots p\}$
\par So in a sense two classes of integrable systems can be manufactured. To be more explicit we consider Kaup-Newell system;
for which
\begin{eqnarray}
U_k&=&-2i\alpha(2)\delta_{1k}+q_k\beta_1(1)+r_k\beta_2(1)\nonumber\\
\end{eqnarray}
\begin{eqnarray}
V_k&=&\sum_{l}[A_k^l\alpha(-2l)+B_{k}^l\beta_1(-2l-1)+C_{k}^l\beta_2(-2l-1)]\nonumber\\
\end{eqnarray}
where $\{k=1\dots p+1\}$ and $\{\alpha_i, \beta_i\}$ are generators of an infinite dimensional Lie algebra
\begin{eqnarray}
[\alpha(\mu),\beta_1(\nu)]=\beta_1(\mu+\nu)\nonumber\\
\end{eqnarray}
\begin{eqnarray}
[\alpha(\mu),\beta_2(\nu)]=-\beta_2(\mu+\nu)\nonumber\\
\end{eqnarray}
\begin{eqnarray}
[\beta_1(\mu),\beta_2(\nu)]=2\alpha(\mu+\nu)
\end{eqnarray}
From the Lax equation equating similar powers of $\lambda$ we get
\begin{eqnarray}
A_{n+1x}^{(N)}&=&2\sum_{\nu=1}^{n+1}[q_{\nu}C_{n-\nu+2}]\nonumber\\
B_{n+1x}^{(N)}&=&-2iB_{n+1}^{(N+1)}-\sum_{\nu=1}^{n+1}q_{\nu} A_{n-\nu+2}\nonumber\\
C_{n+1x}^{(N)}&=&2iC_{n+1}^{(N+1)}+\sum_{\nu=1}^{n+1}r_{\nu}A_{n-\nu+2}
\end{eqnarray}
for $\{n=1\dots p\}$
This leads  to a recursion relation relation of the following form;
\begin{eqnarray}
B_{n+1}^{(N+1)}=\frac{1}{2}\sum_{\nu=1}^{n+1}[i\delta_{\nu,n+1}B_{n+1x}^{(N)}+\sum_{\sigma=1}^{\nu}\Theta_{(\nu\sigma)}^{qr}B_{\sigma x}^{(N)}]\nonumber\\
+\frac{1}{2}\sum_{\nu=1}^{n+1}\sum_{\sigma=1}^{\nu}\Theta_{(\nu\sigma)}^{qq}C_{\sigma x}^{(N)}
\end{eqnarray}
\begin{eqnarray}
C_{n+1}^{(N+1)}=\frac{1}{2}\sum_{\nu=1}^{n+1}[-i\delta_{\nu,n+1}C_{n+1x}^{(N)}+\sum_{\sigma=1}^{\nu}\Theta_{(\nu\sigma)}^{rq}C_{\sigma x}^{(N)}]\nonumber\\
+\frac{1}{2}\sum_{\nu=1}^{n+1}\sum_{\sigma=1}^{\nu}\Theta_{(\nu\sigma)}^{rr}B_{\sigma x}^{(N)}
\end{eqnarray}

where $\Theta$ stands for integral operator
\begin{eqnarray}
\Theta_{(\nu\sigma)}^{xy}=X_{n-\nu+2}\partial^{-1}Y_{n-\sigma+2}=\Theta_{n-\nu+2,n-\sigma+2}
\end{eqnarray}
\par In order to get explicit equations from the above we set
\begin{eqnarray}
B_{k+1}=(2\delta_{1,k+1}-1)r_{k+1}\nonumber\\
C_{k+1}=(2\delta_{1,k+1}-1)q_{k+1}
\end{eqnarray}
\begin{eqnarray}
q_{n+1t}=\frac{i}{2}(2\delta_{1,n+1}-1)r_{n+1xx}+\frac{1}{2}\sum_{\nu=1}^{n+1}\sum_{\sigma=1}^{\nu}[q_{n-\nu+2}\partial_{-1}(q_{n-\sigma+2}q_{\sigma x}+r_{n-\sigma+2}r_{\sigma x})(2\delta_{1\sigma}-1)]_x
\end{eqnarray}
\begin{eqnarray}
r_{n+1t}=-\frac{i}{2}(2\delta_{1,n+1}-1)q_{n+1xx}+\frac{1}{2}\sum_{\nu=1}^{n+1}\sum_{\sigma=1}^{\nu}[r_{n-\nu+2}\partial_{-1}(q_{n-\sigma+2}q_{\sigma x}+r_{n-\sigma+2}r_{\sigma x})(2\delta_{1\sigma}-1)]_x
\end{eqnarray}
where $\{n=1\dots p\}$
\par For example when $p=2$ we get
\begin{eqnarray}
q_{1t}=\frac{i}{2}r_{1xx}+\frac{1}{4}[q_1(r_1^2+q_1^2)]_x
\end{eqnarray}
\begin{eqnarray}
r_{1t}=-\frac{i}{2}q_{1xx}+\frac{1}{4}[r_1(r_1^2+q_1^2)]_x
\end{eqnarray}
\begin{eqnarray}
q_{2t}=-\frac{i}{2}r_{2xx}+\frac{1}{2}[q_1\{\partial^{-1}(q_2q_{1x}-q_1q_{2x}+r_2r_{1x}-r_1r_{2x})\}+\frac{1}{2}q_2(q_1^2+r_1^2)]_x
\end{eqnarray}
\begin{eqnarray}
r_{2t}=\frac{i}{2}q_{2xx}+\frac{1}{2}[r_1\{\partial^{-1}(q_2q_{1x}-q_1q_{2x}+r_2r_{1x}-r_1r_{2x})\}+\frac{1}{2}r_2(q_1^2+r_1^2)]_x
\end{eqnarray}
\begin{eqnarray}
q_{3t}=-\frac{i}{2}r_{3xx}+\frac{1}{2}[q_1\{\partial^{-1}(q_3q_{1x}-q_1q_{3x}+r_3r_{1x}-r_1r_{3x})\}+\frac{1}{2}q_3(q_1^2+r_1^2)]_x\nonumber\\
+\frac{1}{2}[q_2\{\partial^{-1}(q_2q_{1x}-q_1q_{2x}+r_2r_{1x}-r_1r_{2x})\}-\frac{1}{2}q_1(q_2^2+r_2^2)]_x
\end{eqnarray}
\begin{eqnarray}
r_{3t}=-\frac{i}{2}q_{3xx}+\frac{1}{2}[r_1\{\partial^{-1}(q_3q_{1x}-q_1q_{3x}+r_3r_{1x}-r_1r_{3x})\}+\frac{1}{2}r_3(q_1^2+r_1^2)]_x\nonumber\\
+\frac{1}{2}[r_2\{\partial^{-1}(q_2q_{1x}-q_1q_{2x}+r_2r_{1x}-r_1r_{2x})\}-\frac{1}{2}r_1(q_2^2+r_2^2)]_x
\end{eqnarray}
\subsubsection{Hamiltonian Structure}
\par From the inception of study of integrable systems it has been observed that they can be deduced from a Hamiltonian and a symplectic operator. Actually many of them are found to be bi-Hamiltonian. An elegant method for the derivation of the Hamiltonian form was proposed by Tu. In this method a Lie algebra and its corresponding loop algebra $\tilde{G}=G\bigotimes C(\lambda, \lambda^{-1})$ is a Laurent polynomial in $\lambda$. Then one introduces the isospectral problem $\psi_x=U\psi$ and $\psi_t=V\psi$ where $U=R+\sum_{i=1}^p u_ie_i$ where $R$ stands for a pseudo-regular element in $\tilde G$ and $\{e_i\mid i=1\dots p\}$ is a basis in $\tilde G$. $\{u_i(x,t)\mid i=1\dots p\}$ are functions called potential functions. One usually assign a degree to every element $\deg(X\bigotimes \lambda^n)=n$, $X\in G$, $\deg(R)=\alpha$ and $\deg(e_i)=\varepsilon_i$. If $\alpha\geq\epsilon$ then one can solve
\begin{eqnarray}
V_x=[U, V]
\end{eqnarray}
for $V$ as done in previous section. The next step is to choose a modified term $\Delta_n\in\tilde{G}$ so that
\begin{eqnarray}
-V_x^{(n)}+[U, V^{(n)}]=-V_{+x}^{(n)}+[U, V_{+}^{(n)}]-(\Delta_n)_x
\end{eqnarray}
where
\begin{eqnarray}
-V_{+}^{(n)}=(\lambda_nV)_{+}
\end{eqnarray}
is the positive part of $\lambda^n V$.
\par The forth step is to  use the full Lax equation to get the nonlinear system with corresponding $V_n$. Finally we seek the Hamiltonian structure by using the trace identity
\begin{eqnarray}
\frac{\delta}{\delta u}\langle{V,\frac{\partial U}{\partial \lambda}}\rangle=\left(\lambda^{-\gamma}\frac{\partial}{\partial\lambda}\lambda^{\gamma}\right)\langle{V,\frac{\partial U}{\partial u_i}}\rangle
\end{eqnarray}
in order to determine the value of $\gamma$.

Here define  a form of scalar product. Our two main assumptions are
\par (i) $N$ is strictly upper triangular of order $p+1$ with
\begin{eqnarray}
N_{ij}=0 \forall i\geq j\nonumber\\
N_{ij}=1 \forall < j
\end{eqnarray}
\par (ii) If $P$ and $Q$ are defined respectively as
\begin{eqnarray}
P=M\otimes N\nonumber\\
Q=A\otimes B
\end{eqnarray}
with the inner product
\begin{eqnarray}
\langle P,Q \rangle=tr(M^TA\otimes NB)
\end{eqnarray}
where $M^T$ is the transpose of $M$.
\par Considering
\begin{eqnarray}
d(A\otimes B)=dA\otimes B + A\otimes dB
\end{eqnarray}
we get
\begin{eqnarray}
\frac{\partial U}{\partial\lambda}=\textbf{1}\otimes\frac{\partial U_1}{\partial\lambda}\nonumber\\
\sum_{j=1}^{p}N^j\otimes\frac{\partial U_{j+1}}{\partial\lambda}\nonumber\\
\frac{\partial U}{\partial q_{m+1}}=N^m\otimes\frac{\partial U_{m+1}}{\partial q_{m+1}}\nonumber\\
\frac{\partial U}{\partial r_{m+1}}=N^m\otimes\frac{\partial U_{m+1}}{\partial r_{m+1}}
\end{eqnarray}
So we get
\begin{eqnarray}
\langle{V,\frac{\partial U}{\partial \lambda}}\rangle=(p+1)[-4i\lambda A_1+C_1q_1+B_1r_1]\nonumber\\
+\sum_{j+1}^{p}\sum_{k+1}^{p}\omega_{jk}(C_{j+1}q_{k+1}+B_{j+1}r_{k+1})
\end{eqnarray}
where $\omega_{jk}$ stands for
$\omega_{jk}=\frac{1}{2}(p-j+1)(p-j+2)$ for $j\geq k$ and $\omega_{jk}=\frac{1}{2}(p-k+1)(p-k+2)$ for $j\leq k$. Obviously $\omega_{jk}=\omega_{kj}$.
Hence we get
\begin{eqnarray}
\langle{V,\frac{\partial U}{\partial q_{l+1}}}\rangle&=&\sum_{j=1}^{p}\lambda\omega_{jl}C_{j+1}\nonumber\\
\langle{V,\frac{\partial U}{\partial r_{l+1}}}\rangle&=&\sum_{j=1}^{p}\lambda\omega_{jl}B_{j+1}
\end{eqnarray}
Now using the fact that $\omega_{jl}$ is symmetric and the trace identity
\begin{eqnarray}
\frac{\delta}{\delta q_{l+1}}\langle{V,\frac{\partial U}{\partial\lambda}}\rangle=\lambda^{-\gamma}\frac{\partial}{\partial\lambda}\lambda^{\gamma}\langle{V,\frac{\partial U}{\partial q_{l+1}}}\rangle
\end{eqnarray}
for $l=1\dots p$ we get $\gamma=-2$.
So, one arrives at the Hamiltonian
\begin{eqnarray}
H_{M}=\frac{1}{2m+2}[(p+1)(4iA_1^{(M+1)}-C_1^{(M)}q_1-B_1^{(M)}r_1)\nonumber\\
-\sum_{j=1}^p\sum_{k=1}^p\omega_{jk}(C_{j+1}^{(M)}q_{k+1}+B_{j+1}^{(M)}r_{k+1})]
\end{eqnarray}
which ultimately leads to Liouville integrability of our system; which has the variational form
\begin{eqnarray}
(q_{1t}, r_{1t}\dots q_{p+1t}, r_{p+1t})^t=J(\frac{\delta}{\delta q_1}, \frac{\delta}{\delta r_1}\dots \frac{\delta}{\delta q_{p+1}}, \frac{\delta}{\delta r_{p+1}})^tH_M
\end{eqnarray}
Here $J$ is the symplectic operator equal to $W^{-1}\partial$ and $W$ takes the following form
\begin{eqnarray}
W= \left(\begin{array}{cccccccccc}
   0 & -1 & 0 & 0 & 0 & 0 & 0 & 0\ldots& 0 & 0\\
   -1 & 0 & 0 & 0 & 0 & 0 & 0 & 0\ldots& 0 & 0\\
   0 & 0 & 0 & -\omega_{12} & 0 & -\omega_{22} & 0 & -\omega_{32}\ldots & 0 & -\omega_{p2}\\
   0 & 0 & -\omega_{12} & 0  & -\omega_{22} & 0 & -\omega_{32} & 0\ldots &  -\omega_{p2} & 0\\
   0 & 0 & 0 & -\omega_{13} & 0 & -\omega_{23} & 0 & -\omega_{33}\ldots & 0 & -\omega_{p3}\\
   0 & 0 & -\omega_{13} & 0  & -\omega_{23} & 0 & -\omega_{33} & 0\ldots &  -\omega_{p3} & 0\\
\vdots&\vdots&\vdots&\vdots&\vdots&\vdots&\vdots&\vdots&\vdots&\vdots\\
   0 & 0 & 0 & -\omega_{1p} & 0 & -\omega_{2p} & 0 & -\omega_{3p}\ldots & 0 & -\omega_{pp}\\
   0 & 0 & -\omega_{1p} & 0  & -\omega_{2p} & 0 & -\omega_{3p} & 0\ldots &  -\omega_{pp} & 0
 \end{array} \right)
\end{eqnarray}
\subsection{Coupling by Hadamard matrix}
As we described before, the coupled set of integrable systems can also be generated with the help of Hadamard matrices. A Hadamrd matrix is defined by an $n\times n$ matrix $\Gamma$ such that
\begin{eqnarray}
\Gamma^T\Gamma=n
\end{eqnarray}
with mutually orthogonal rows and columns.
Here we have taken a $2\times 2$ Hadamard matrix
\begin{eqnarray}
\Gamma^{(1)}= \left(\begin{array}{cc}
   1 & -1 \\
   1 & 1
    \end{array} \right)
\end{eqnarray}
 along with other matrices like $\Gamma^{(2)}={\Gamma^{(1)}}^T$ and $\Gamma^{(3)}=\frac{1}{2}(\Gamma^{(1)}-\Gamma^{(2)})$
\par With these we construct
\begin{eqnarray}
U=\textbf{1}\otimes U_1+\sum_{j=1}^3\Gamma^{(j)}\otimes U_{j+1}\nonumber\\
V=\textbf{1}\otimes V_1+\sum_{j=1}^3\Gamma^{(j)}\otimes V_{j+1}
\end{eqnarray}
The zero curvature equation leads to
\begin{eqnarray}
U_{t}-V_{x}+[U,V]&=&0
\end{eqnarray}
Hence
\begin{eqnarray}
U_{1t}-V_{1x}+[U_1,V_1]+2[U_2,V_3]+2[U_3,V_2]-[U_4,V_4]=0\nonumber\\
U_{2t}-V_{2x}+[U_1,V_2]+[U_2,V_1]+[U_3,V_4]-[U_4,V_3]=0\nonumber\\
U_{3t}-V_{3x}+[U_1,V_3]-[U_2,V_4]+[U_3,V_1]-[U_4,V_2]=0\nonumber\\
U_{4t}-V_{4x}+[U_1,V_4]+2[U_2,V_2]-2[U_3,V_3]+[U_4,V_1]=0
\end{eqnarray}
Taking
\begin{eqnarray}
U_k=-2\delta_{1k}\alpha(1)+q_k\beta_1(0)+r_k\beta_2(0)\nonumber\\
V_k=\sum[A_k^{(j)}\alpha(-j)+B_k^{(j)}\beta_1(-j)+C_k^{(j)}\beta_2(-j)]
\end{eqnarray}
where $k=1, 2, 3, 4$.
\par Equating the coefficients of $\alpha,\beta_1,\beta_2$ we get
\begin{eqnarray}
A_{1x}^{(m)}&=&2q_1C_1^{(m)}-2r_1B_1^{(m)}+2(2q_2C_3^{(m)}-2r_2B_3^{(m)})+4q_3C_2^{(m)}-4r_3B_2^{(m)}-(2q_4C_4^{(m)}-2r_4B_4^{(m)})\nonumber\\
B_{1x}^{(m)}&=&-2B_1^{(m+1)}-q_1A_1^{(m)}-2q_2A_3^{(m)}-2q_3A_2^{(m)}+q_4A_4^{(m)}\nonumber\\
C_{1x}^{(m)}&=&2C_1^{(m+1)}+r_1A_1^{(m)}+2r_2A_3^{(m)}+2r_3A_2^{(m)}-r_4A_4^{(m)}
\end{eqnarray}
along with
\begin{eqnarray}
A_{2x}^{(m)}&=&2q_1C_2^{(m)}-2r_1B_2^{(m)}+2(2q_2C_1^{(m)}-2r_2B_1^{(m)})+2q_3C_4^{(m)}-2r_3B_4^{(m)}+(2q_4C_3^{(m)}-2r_4B_3^{(m)})\nonumber\\
B_{1x}^{(m)}&=&-2B_2^{(m+1)}-q_1A_2^{(m)}-q_2A_1^{(m)}-q_3A_4^{(m)}-q_4A_3^{(m)}\nonumber\\
C_{1x}^{(m)}&=&2C_2^{(m+1)}+r_1A_2^{(m)}+r_2A_1^{(m)}+r_3A_4^{(m)}+r_4A_3^{(m)}\nonumber\\
\end{eqnarray}
\begin{eqnarray}
A_{3x}^{(m)}&=&2q_1C_3^{(m)}-2r_1B_3^{(m)}+(2q_3C_1^{(m)}-2r_3B_1^{(m)})-2q_2C_4^{(m)}+2r_4B_4^{(m)}-(2q_4C_2^{(m)}+2r_4B_2^{(m)})\nonumber\\
B_{3x}^{(m)}&=&-2B_3^{(m+1)}-q_1A_3^{(m)}-q_3A_1^{(m)}+q_2A_4^{(m)}+q_4A_2^{(m)}\nonumber\\
C_{3x}^{(m)}&=&2C_3^{(m+1)}+r_1A_3^{(m)}+r_3A_1^{(m)}-r_2A_4^{(m)}-r_4A_2^{(m)}
\end{eqnarray}
\begin{eqnarray}
A_{4x}^{(m)}&=&2q_1C_4^{(m)}-2r_1B_4^{(m)}+2(2q_2C_2^{(m)}-2r_2B_2^{(m)})-4q_3C_3^{(m)}+4r_3B_3^{(m)}+(2q_4C_1^{(m)}-2r_4B_1^{(m)})\nonumber\\
B_{4x}^{(m)}&=&-2B_4^{(m+1)}-q_1A_4^{(m)}-2q_2A_2^{(m)}+2q_3A_3^{(m)}-q_4A_1^{(m)}\nonumber\\
C_{4x}^{(m)}&=&2C_4^{(m+1)}+r_1A_4^{(m)}+2r_2A_2^{(m)}-2r_3A_3^{(m)}+r_4A_1^{(m)}
\end{eqnarray}
leading to a recursion matrix
\begin{eqnarray}
\Omega^{(i)}=(B_1^{(i)},C_1^{(i)},B_2^{(i)},C_2^{(i)},B_3^{(i)},C_3^{(i)},B_4^{(i)}C_4^{(i)})
\end{eqnarray}

\begin{eqnarray}
\Omega^{(m+1)}=M\Omega^{(m)}
\end{eqnarray}
The explicit forms of the elements of recursion matrix $M$  is enlisted in the appendix.
With the help of this recursion relation we can get the following nonlinear equations.
\begin{eqnarray}
q_{1t}=-\frac{1}{2}q_{1xx}+r_1(q_1^2-q_4^2)+2r_4(q_3^2-q_2^2)+4(q_1q_2r_3+q_1q_3r_2+q_2q_3r_1-q_4q_2r_2+q_3q_4r_3)-2q_1q_4r_4
\end{eqnarray}
\begin{eqnarray}
r_{1t}=\frac{1}{2}r_{1xx}+q_1(r_4^2-r_1^2)+2q_4(r_2^2-r_3^2)+4(q_2r_2r_4-r_1q_2r_3-r_2q_3r_1-q_1r_2r_3-q_3r_4r_3)+2r_1q_4r_4
\end{eqnarray}
\begin{eqnarray}
q_{2t}=-\frac{1}{2}q_{2xx}+r_2(q_1^2-q_4^2)+2r_3(q_2^2-q_3^2)+2(q_1q_2r_1+q_1q_3r_4+q_1q_4r_3-q_4q_2r_4+q_3q_4r_1)+4q_3q_2r_2
\end{eqnarray}
\begin{eqnarray}
r_{2t}=\frac{1}{2}r_{2xx}+q_2(r_4^2-r_1^2)+2q_3(r_3^2-r_2^2)+2(q_4r_2r_4-q_1r_1r_2-r_4q_3r_1-q_4r_3r_1-q_1r_4r_3)+4q_2r_32r_3
\end{eqnarray}
\begin{eqnarray}
q_{3t}=-\frac{1}{2}q_{3xx}+r_3(q_1^2-q_4^2)+2r_2(q_3^2-q_2^2)+2(q_1q_3r_1-q_1q_2r_4-q_1q_4r_2-q_4q_3r_4-q_2q_4r_1)+4q_2q_3r_3
\end{eqnarray}
\begin{eqnarray}
r_{3t}=\frac{1}{2}r_{3xx}+r_1(r_4^2-r_1^2)+2q_2(r_2^2-r_3^2)+2(r_1q_2r_4-q_1r_3r_1+r_2q_4r_1+q_4r_3r_4+q_1r_4r_2)-4q_3r_2r_3
\end{eqnarray}
\begin{eqnarray}
q_{4t}=-\frac{1}{2}q_{4xx}+r_4(q_1^2-q_4^2)+2r_1(q_2^2-q_3^2)+4(q_1q_2r_2-q_1q_3r_3+q_2q_3r_4+q_4q_2r_3+q_2q_4r_2)+2q_1q_4r_1
\end{eqnarray}
\begin{eqnarray}
r_{4t}=\frac{1}{2}r_{4xx}+q_4(r_4^2-r_1^2)+2q_1(r_3^2-r_2^2)+4(-q_2r_2r_1+r_1q_3r_3-r_2q_3r_4-q_4r_3r_2-q_2r_4r_3)-2q_1r_1r_4
\end{eqnarray}
\subsubsection{Hamiltonian Structure}
We can construct various relevant traces as inner products we find
\begin{eqnarray}
\langle{V,\frac{\partial U}{\partial \lambda}}\rangle=-4(A_1+A_2+A_3)\nonumber\\
\langle{V,\frac{\partial U}{\partial q_1}}\rangle=2(C_1+C_2+C_3)\nonumber\\
\langle{V,\frac{\partial U}{\partial q_2}}\rangle=2(C_1+2C_3-C_4)\nonumber\\
\langle{V,\frac{\partial U}{\partial q_3}}\rangle=2(C_1+2C_2-C_4)\nonumber\\
\langle{V,\frac{\partial U}{\partial q_4}}\rangle=2(C_3-C_2+C_4)
\end{eqnarray}
Similar expressions can be obtained for $\langle{V,\frac{\partial U}{\partial r_i}}\rangle$ for $i=1, 2, 3, 4$. Using trace identity as usual we obtain $\gamma=-2$ leading to an expression Hamiltonians
\begin{eqnarray}
H_m=\frac{4}{m+2}\sum_{j=1}^3A_j^{(m+1)}
\end{eqnarray}
The so called Liouville integrability can be ensured by observing the following equation of motion
\begin{eqnarray}
J(\frac{\delta}{\delta q_i},\frac{\delta}{\delta r_i}\mid i=1,2,3,4)^tH_m=(q_{it},r_{it}\mid i=1,2,3,4)
\end{eqnarray}
Where $J$ is a symplectic operator given by $J=W^{-1}$ with
\begin{eqnarray}
J= \left(\begin{array}{cccccccc}
   0 & 1 & 0 & 1 & 0 & 1 & 0 & 0\\
   -1 & 0 & -1 & 0 & -1 & 0 & 0 & 0\\
   0 & 1 & 0 & 0 & 0 & 2 & 0 & -1\\
   -1 & 0 & 0 & 0 & -2 & 0 & 1 & 0\\
   0 & 1 & 0 & 2 & 0 & 0 & 0 & -1\\
   -1 & 0 & -2 & 0 & 0 & 0 & 1 & 0\\
   0 & 0 & 0 & -1 & 0 & 1 & 0 & -1\\
   0 & 0 & 1 & 0 & -1 & 0 & 1 & 0
   \end{array} \right)
\end{eqnarray}
\subsection{Coupling by idempotent matrices}
A matrix $P$ is said to be idempotent iff $P^2=P$. For our present purpose we take a collection of $n$ idempotent matrices with the following structure
\begin{eqnarray}
(P_j)_{\mu\nu}=1 \forall \mu=\nu\leq j\nonumber\\
(P_j)_{\mu\nu}=0 otherwise
\end{eqnarray}
Defining the time and space parts of the Lax operator
\begin{eqnarray}
U=\sum_{1}^n P_j\otimes U_j\nonumber\\
V=\sum_{1}^n P_j\otimes V_j
\end{eqnarray}
Hence the zero curvature equations involving $U$ and $V$ leads to
\begin{eqnarray}
U_{lt}-V_{lt}+[U_l, \sum_{m=0}^{n-1}V_{m+l}]+[\sum_{m=0}^{n-l-1}U_{l+m+1}, V_l]=0\nonumber\\
\end{eqnarray}
where $l=1\dots n$
The generalized recursion relations can be written as
\begin{eqnarray}
A_{lx}^{(j)}&=&2q_l\sum_{m=0}^{n-l}C^{(j)}_{l+m}-2r_l\sum_{m=0}^{n-l}B_{l+m}^{(j)}+2\sum_{m=0}^{n-l-1}q_{l+m+1}C_l^{(j)}
-2\sum_{m=0}^{n-l-1}r_{l+m+1}B_l^{(j)}\nonumber\\
B_{lx}^{(j)}&=&-q_l\sum_{m=0}^{n-l}A^{(j)}_{l+m}-A_l^{(j)}\sum_{m=0}^{n-l-1}q_{l+m+1}-2B_{l}^{(j+1)}\nonumber\\
C_{lx}^{(j)}&=&r_l\sum_{m=0}^{n-l}A^{(j)}_{l+m}+A_l^{(j)}\sum_{m=0}^{n-l-1}r_{l+m+1}+2C_{l}^{(j+1)}
\end{eqnarray}
Again choosing
\begin{eqnarray}
U_l&=&-2\alpha(1)\delta_{ln}+q_l\beta_1(0)+r_l\beta_2(0)\nonumber\\
V_l&=&\sum[A_l^{(j)}\alpha(-j)+B_l^{(j)}\beta_1(-j)+C_l^{(j)}\beta_2(-j)]
\end{eqnarray}
which gives rise to the recursion relation for $\{B_j, C_j\mid j=1\dots n\}$. For example
\begin{eqnarray}
B_l^{(j+1)}=-\frac{1}{2}B_{lx}^{(j)}-\sum_{m=0}^{n-l}\sum_{\mu=0}^{n-l-m}(\Theta^{qq}_{l, l+m}C_{l+\mu+m}^{(j)}-\Theta^{qr}_{l, l+m}B_{l+\mu+m}^{(j)})\nonumber\\
-\sum_{m=0}^{n-l}\sum_{\mu=0}^{n-l-m-1}(\Theta^{qq}_{l, l+m+\mu+1}C_{l+\mu+m}^{(j)}-\Theta^{qr}_{l, l+m+\mu+1}B_{l+\mu+m}^{(j)})\nonumber\\
-\sum_{m=0}^{n-l-1}\sum_{\mu=0}^{n-l}(\Theta^{qq}_{l+m+1,l}C_{l+\mu}^{(j)}-\Theta^{qr}_{l+m+1,l}B_{l+\mu}^{(j)})\nonumber\\
-\sum_{m=0}^{n-l-1}\sum_{\mu=0}^{n-l-1}(\Theta^{qq}_{l+m+1, l+\mu+1}C_{l}^{(j)}-\Theta^{qr}_{l+m+1, l+\mu+1}B_{l}^{(j)})\nonumber\\
\end{eqnarray}
Similar recursion relation for $C_l^{(j+1)}$ can be obtained by replacing $B_l^{(j)}$ with $-C_l^{(j)}$, $B_l^{(j+1)}$ with $C_l^{(j+1)}$ and $\Theta^{qq}_{\mu\nu}$ with $\Theta^{rq}_{\mu\nu}$ where as $\Theta^{qr}_{\mu\nu}$ with $\Theta^{rr}_{\mu\nu}$. Here $\Theta_{\mu\nu}^{xy}$ is defined as
\begin{eqnarray}
\Theta_{\mu\nu}^{xy}=X_{\mu}\partial^{-1}Y_{\nu}
\end{eqnarray}
The corresponding nonlinear equations are
\begin{eqnarray}
q_{lt}=-2B_{l}^{(j+1)}\nonumber\\
r_{lt}=2C_{l}^{(j+1)}
\end{eqnarray}
where $\{l=1\dots n\}$ and $j$ being the index of hierarchy.
Now letting
\begin{eqnarray}
B_l^{(0)}=q_l\nonumber\\
C_l^{(0)}=r_l
\end{eqnarray}
and $A_l=0$ we can obtain a specific hierarchy
\subsubsection{Hamiltonian structure}
\begin{eqnarray}
\langle{V,\frac{\partial U}{\partial \lambda}}\rangle&=&-2\sum_{j=1}^n jA_j\nonumber\\
\langle{V,\frac{\partial U}{\partial q_{\nu}}}\rangle&=&\sum_{j=1}^{\nu-1}jC_j+\nu\sum_{k=\nu}^nC_k\nonumber\\
\langle{V,\frac{\partial U}{\partial r_{\nu}}}\rangle&=&\sum_{j=1}^{\nu-1}jB_j+\nu\sum_{k=\nu}^nB_k\nonumber\\
\end{eqnarray}
where $\nu=1\dots n$
along with
\begin{eqnarray}
H_m=\frac{2}{m+2}\sum_{j=1}^njA_j^{(m+1)}
\end{eqnarray}
and the corresponding Liouville integrability can be demonstrated as
\begin{eqnarray}
(q_{lt}, r_{lt}\mid l=1\dots n)^T=J(\frac{\delta}{\delta q_l}, \frac{\delta}{\delta r_l}\mid l=1\dots n)^TH_m
\end{eqnarray}
with
\begin{eqnarray}
J=W^{-1}
\end{eqnarray}
and

\begin{eqnarray}
W= \frac{1}{2}\left(\begin{array}{cccccccccccc}
   0 & 1 & 0 & 1 & 0 & 1 & 0 & 1\ldots& 0 & 1 & 0 & 1\\
   -1 & 0 & -1 & 0 & -1 & 0 & -1 & 0\ldots& -1 & 0 & -1 & 0\\
   0 & 1 & 0 & 2 & 0 & 1 & 0 & 1\ldots & 0 & 2 & 0 & 2 \\
   -1 & 0 & -2 & 0  & -2 & 0 & -2 & 0\ldots &  -2 & 0 & -2 & 0\\
   0 & 1 & 0 & 2 & 0 & 3 & 0 & 3\ldots & 0 & 3 & 0 & 3\\
   -1 & 0 & -2 & 0  & -3 & 0 & -3 & 0\ldots &  -3 & 0 & -3 & 0\\
\vdots&\vdots&\vdots&\vdots&\vdots&\vdots&\vdots&\vdots&\vdots&\vdots&\vdots&\vdots\\
   0 & 1 & 0 & 2 & 0 & 3 & 0 & 4\ldots & 0 & \nu & 0 & \nu\\
   -1 & 0 & -2 & 0  & -3 & 0 & -4 & 0\ldots &  -\nu & 0 & -\nu & 0\\
   \vdots&\vdots&\vdots&\vdots&\vdots&\vdots&\vdots&\vdots&\vdots&\vdots&\vdots&\vdots\\
   0 & 1 & 0 & 2 & 0 & 3 & 0 & 4\ldots & 0 & n & 0 & n\\
   -1 & 0 & -2 & 0  & -3 & 0 & -4 & 0\ldots &  -n & 0 & -n & 0
 \end{array} \right)
\end{eqnarray}
\subsection{Coupling by K-idempotent matrix}
A matrix $P$ is said to be $K$-idempotent if there exists $K$ such that $KP^2K=P$, $K$ being an associated permutation matrix. The basic properties of such $P$ are
\begin{eqnarray}
KPK&=&P^2\nonumber\\
KP&=&P^2K, KP^2=PK\nonumber\\
P^3K&=&KP^3, KP^3K=P^3\nonumber\\
P^3&=&(KP)^2=(PK)^2
\end{eqnarray}
Following table decodes product table of powers of $P$

\begin{center}
\begin{tabular}{c|c|c|c}
& $P$ & $P^2$ & $P^3$\\
\hline
$P$ & $P^2$ & $P^3$ & $P$\\
\hline
$P^2$ & $P^3$ & $P$ & $P^2$\\
\hline
$P^3$ & $P$ & $P^2$ & $P^3$
\end{tabular}
\end{center}

\par There are many choices of $P$. We have chosen $P$ as
\begin{eqnarray}
P&=&-\frac{1}{2}\textbf{1}+\frac{i}{2}\sqrt{7}\sigma_3+\sigma_2\nonumber\\
K&=&\sigma_1
\end{eqnarray}
with
\begin{eqnarray}
\sigma_1= \left(\begin{array}{cc}
   0 & 1 \\
   1 & 0 \\
 \end{array} \right)
\end{eqnarray}
\begin{eqnarray}
\sigma_2= \left(\begin{array}{cc}
   0 & -i \\
   i & 0 \\
 \end{array} \right)
\end{eqnarray}
\begin{eqnarray}
\sigma_3= \left(\begin{array}{cc}
   1 & 0 \\
   0 & -1 \\
 \end{array} \right)
\end{eqnarray}
For time and space part of the relevant Lax operator we take
\begin{eqnarray}
U=\sum_{j=1}^3P^j\otimes U_{4-j}\nonumber\\
V=\sum_{j=1}^3P^{j}\otimes V_{4-j}\nonumber\\
\end{eqnarray}
Using Lax condition
\begin{eqnarray}
U_t-V_x+[U,V]=0
\end{eqnarray}
\begin{eqnarray}
U_{1t}-V_{1x}+[U_1,V_1]+[U_2,V_3]+[U_3,V_2]=0\nonumber\\
U_{2t}-V_{2x}+[U_1,V_2]+[U_2,V_1]+[U_3,V_3]=0\nonumber\\
U_{3t}-V_{3x}+[U_1,V_3]+[U_2,V_2]+[U_3,V_1]=0\nonumber\\
\end{eqnarray}
As a choice we take
\begin{eqnarray}
U_k&=&-2\alpha(1)\delta_{1k}+q_k\beta_1(0)+r_k\beta_2(0)\nonumber\\
V_{k}&=&\sum_j[A_k^{(j)}\alpha(-j)+B_k^{(j)}\beta_1(-j)+C_k^{(j)}\beta_2(-j)]
\end{eqnarray}
The consistency condition leads to the recursion relation
\begin{eqnarray}
A_{1x}^{(m)}&=&2q_1C_1^{(m)}-2r_1B_1^{(m)}+2q_2C_3^{(m)}-2r_2B_3^{(m)}+2q_3C_2^{(m)}-2r_3B_2^{(m)}\nonumber\\
B_{1x}^{(m)}&=&-2B_1^{(m+1)}-q_1A_1^{(m)}-q_2A_3^{(m)}-q_3A_2^{(m)}\nonumber\\
C_{1x}^{(m)}&=&2C_1^{m+1}+r_1A_1^{(m)}+r_2A_3^{(m)}+r_3A_2^{(m)}
\end{eqnarray}
with similar relations for $A_2,B_2, C_2$ and $A_3, B_3, C_3$.
In matrix form we can write
\begin{eqnarray}
\Pi_{m+1}=N \Pi_m
\end{eqnarray}
where
\begin{eqnarray}
\Pi_{m}=(B_j^{(m)},C_j^{(m)}\mid j=1,2,3)
\end{eqnarray}
\subsubsection{Hamiltonian structure}
For the Hamiltonian structure of such equations we note that $Tr P=-1=TrP^2$ and $TrP^3=2$.
\begin{eqnarray}
\langle{V,\frac{\partial U}{\partial \lambda}}\rangle&=&-2(2A_1-A_2-A_3)\nonumber\\
\langle{V,\frac{\partial U}{\partial q_{1}}}\rangle&=&2C_1-C_2-C_3\nonumber\\
\langle{V,\frac{\partial U}{\partial q_{2}}}\rangle&=&2C_3-C_1-C_2\nonumber\\
\langle{V,\frac{\partial U}{\partial q_{3}}}\rangle&=&2C_2-C_1-C_3\nonumber\\
\langle{V,\frac{\partial U}{\partial r_{1}}}\rangle&=&2B_1-B_2-B_3\nonumber\\
\langle{V,\frac{\partial U}{\partial r_{2}}}\rangle&=&2B_3-B_1-B_2\nonumber\\
\langle{V,\frac{\partial U}{\partial r_{3}}}\rangle&=&2B_2-B_1-B_3\nonumber\\
\end{eqnarray}
Applying trace identity the hamiltonian becomes
\begin{eqnarray}
H_m=\frac{2}{m+2}(2A_1^{(m+1)}-A_2^{(m+1)}-A_3^{(m+1)})
\end{eqnarray}
so that the nonlinear system can be written as
\begin{eqnarray}
(q_{lt}, r_{lt}\mid l=1\dots 3)^T=J(\frac{\delta}{\delta q_l}, \frac{\delta}{\delta r_l}\mid l=1\dots 3)^TH_m
\end{eqnarray}
where $J=W^{-1}$ and
\begin{eqnarray}
W= \frac{1}{2}\left(\begin{array}{cccccc}
   0 & 2 & 0 & -1 & 0 & -1\\
   -2 & 0 & 1 & 0 & 1 & 0\\
   0 & -1 & 0 & -1 & 0 & 2\\
   1 & 0 & 1 & 0 & -2 & 0\\
   0 & -1 & 0 & 2 & 0 & -1\\
   1 & 0 & -2 & 0 & 1 & 0
   \end{array} \right)
\end{eqnarray}

\section{Conclusion}
The novelty of  the method discussed above in various cases lies in the choice of matrices with which various loop algebraic elements are combined via Kronecker product. The coupling and nonlinearity thus obtained raises apprehension of the existence of various other such matrices leading to several other nonlinear coupled equations not yet known. Those equations are also expected to carry several manifestly important signatures of algebraic properties of the matrices used. Though the fundamental zero curvature equation is always at the starting point of our formulation the Liouville integrability through Hamiltonian structure of all such equation is almost ensured to within a choice of inner product. Possible extension of this method can be suggested in (2+1) or even in (n+1) dimensions and for both continuous and discrete types of flow. An associated problem in non-isospectral regime can also be envisaged.

\section{Appendix}
\par I. Elements of matrix $M$ in equation(51)
\begin{eqnarray}
M_{11}&=&-\frac{1}{2}\partial+\Theta^{qr}_{11}+2(\Theta^{qr}_{23}+\Theta^{qr}_{32})-\Theta^{qr}_{44}\nonumber\\
M_{12}&=&-\Theta^{qq}_{11}-2(\Theta^{qq}_{23}+\Theta^{qq}_{32})+\Theta^{qq}_{44}\nonumber\\
M_{13}&=&2(\Theta^{qr}_{13}+\Theta^{qr}_{31}-\Theta^{qr}_{24}-\Theta^{qr}_{42})\nonumber\\
M_{14}&=&2(\Theta^{qq}_{24}+\Theta^{qq}_{42}-\Theta^{qq}_{13}-\Theta^{qq}_{31})\nonumber\\
M_{15}&=&2(\Theta^{qr}_{12}+\Theta^{qr}_{21}+\Theta^{qr}_{34}+\Theta^{qr}_{43})\nonumber\\
M_{16}&=&-2(\Theta^{qq}_{12}+\Theta^{qq}_{21}+\Theta^{qq}_{34}+\Theta^{qq}_{43})\nonumber\\
M_{17}&=&-\Theta^{qr}_{14}-2(\Theta^{qr}_{22}-\Theta^{qr}_{33})-\Theta^{qr}_{41}\nonumber\\
M_{18}&=&\Theta^{qq}_{14}+2(\Theta^{qq}_{22}-\Theta^{qq}_{33})+\Theta^{qq}_{44}\nonumber\\
M_{21}&=&\Theta^{rr}_{11}+2(\Theta^{rr}_{23}+\Theta^{rr}_{32})-\Theta^{rr}_{44}\nonumber\\
M_{22}&=&\frac{1}{2}\partial-\Theta^{rq}_{11}-2(\Theta^{rq}_{23}+\Theta^{rq}_{32})+\Theta^{rq}_{44}\nonumber\\
M_{23}&=&2(\Theta^{rr}_{13}-\Theta^{rr}_{24}+\Theta^{rr}_{31}-\Theta^{rr}_{42})\nonumber\\
M_{24}&=&-2(\Theta^{rq}_{13}-\Theta^{rq}_{24}+\Theta^{rq}_{31}-\Theta^{rq}_{42})\nonumber\\
M_{25}&=&2(\Theta^{rr}_{12}+\Theta^{rr}_{21}+\Theta^{rr}_{34}+\Theta^{rr}_{43})\nonumber\\
M_{26}&=&-2(\Theta^{rq}_{12}+\Theta^{rq}_{21}+\Theta^{rq}_{34}+\Theta^{rq}_{43})\nonumber\\
M_{27}&=&-\Theta^{rr}_{14}-2(\Theta^{rr}_{22}-\Theta^{rr}_{33})-\Theta^{rr}_{41}\nonumber\\
M_{28}&=&\Theta^{rq}_{14}+2(\Theta^{rq}_{22}-\Theta^{rq}_{33})+\Theta^{rq}_{41}\nonumber\\
M_{31}&=&(\Theta^{qr}_{12}+\Theta^{qr}_{21}+\Theta^{qr}_{34}+\Theta^{qr}_{43})\nonumber\\
M_{32}&=&-(\Theta^{qq}_{12}+\Theta^{qq}_{21}+\Theta^{qq}_{34}+\Theta^{qq}_{43})\nonumber\\
M_{33}&=&-\frac{1}{2}\partial+\Theta^{qr}_{11}+2(\Theta^{qr}_{23}+\Theta^{rq}_{32})-\Theta^{qr}_{44}\nonumber\\
M_{34}&=&-\Theta^{qq}_{11}-2(\Theta^{qq}_{23}+\Theta^{qq}_{32})+\Theta^{qq}_{44}\nonumber\\
M_{35}&=&\Theta^{qr}_{14}+2(\Theta^{qr}_{22}-\Theta^{qr}_{33})+\Theta^{qr}_{41}\nonumber\\
M_{36}&=&-\Theta^{qq}_{14}-2(\Theta^{qq}_{24}-\Theta^{qq}_{33})-\Theta^{qq}_{41}\nonumber\\
M_{37}&=&(\Theta^{qr}_{13}-\Theta^{qr}_{21}+\Theta^{qr}_{31}-\Theta^{qr}_{42})\nonumber\\
M_{38}&=&-(\Theta^{qq}_{13}+\Theta^{qq}_{24}-\Theta^{qq}_{31}+\Theta^{qq}_{42})\nonumber\\
M_{41}&=&(\Theta^{rr}_{12}+\Theta^{rr}_{21}+\Theta^{rr}_{34}+\Theta^{rr}_{43})\nonumber\\
M_{42}&=&-(\Theta^{rq}_{12}+\Theta^{rq}_{21}+\Theta^{rq}_{34}+\Theta^{rq}_{43})\nonumber\\
M_{43}&=&(\Theta^{rr}_{11}+2\Theta^{rr}_{23}+2\Theta^{rr}_{32}-\Theta^{rr}_{44})\nonumber\\
M_{44}&=&\frac{1}{2}\partial-(\Theta^{rq}_{11}-2\Theta^{rq}_{23}-2\Theta^{rq}_{32}+\Theta^{rq}_{44})\nonumber\\
M_{45}&=&(\Theta^{rr}_{14}+2\Theta^{rr}_{22}+2\Theta^{rr}_{33}+\Theta^{rr}_{41})\nonumber\\
M_{46}&=&-(\Theta^{rq}_{14}+2\Theta^{rq}_{22}+2\Theta^{rq}_{33}-\Theta^{rq}_{41})\nonumber\\
M_{47}&=&(\Theta^{rr}_{13}-\Theta^{rr}_{24}+\Theta^{rr}_{31}-\Theta^{rq}_{42})\nonumber\\
M_{48}&=&-(\Theta^{rq}_{13}-\Theta^{rq}_{24}+\Theta^{rq}_{31}-\Theta^{rq}_{42})\nonumber\\
M_{51}&=&(\Theta^{qr}_{13}+\Theta^{qr}_{31}-\Theta^{qr}_{24}-\Theta^{qr}_{42})\nonumber\\
M_{52}&=&(\Theta^{qq}_{24}+\Theta^{qq}_{42}-\Theta^{qq}_{13}-\Theta^{qq}_{31})\nonumber\\
M_{53}&=&(-\Theta^{qr}_{14}+2\Theta^{qr}_{33}-2\Theta^{qr}_{22}-\Theta^{qr}_{41})\nonumber\\
M_{54}&=&(\Theta^{qq}_{14}+\Theta^{qq}_{41}-2\Theta^{qq}_{33}+2\Theta^{qq}_{22})\nonumber\\
M_{55}&=&-\frac{1}{2}\partial+(\Theta^{qr}_{11}+2\Theta^{qr}_{32}+\Theta^{qr}_{23}-\Theta^{qr}_{44})\nonumber\\
M_{56}&=&(-\Theta^{qq}_{11}-2\Theta^{qq}_{32}-2\Theta^{qq}_{23}+\Theta^{qq}_{44})\nonumber\\
M_{57}&=&-(\Theta^{qr}_{12}+\Theta^{qr}_{34}+\Theta^{qr}_{21}+\Theta^{qr}_{43})\nonumber\\
M_{58}&=&(\Theta^{qq}_{12}+\Theta^{qq}_{34}+\Theta^{qq}_{21}+\Theta^{qq}_{43})\nonumber\\
M_{61}&=&(\Theta^{rr}_{13}+\Theta^{rr}_{31}+\Theta^{rr}_{24}+\Theta^{rr}_{42})\nonumber\\
M_{62}&=&(-\Theta^{rq}_{13}-\Theta^{rq}_{31}-\Theta^{rq}_{24}+\Theta^{rq}_{42})\nonumber\\
M_{63}&=&(-\Theta^{rr}_{14}+2\Theta^{rr}_{33}+2\Theta^{qr}_{22}-\Theta^{rr}_{41})\nonumber\\
M_{64}&=&(\Theta^{rq}_{14}-2\Theta^{rq}_{33}-2\Theta^{rq}_{22}+\Theta^{rq}_{41})\nonumber\\
M_{65}&=&(\Theta^{rr}_{11}+2\Theta^{rr}_{32}-2\Theta^{rr}_{23}-\Theta^{rr}_{44})\nonumber\\
M_{66}&=&\frac{1}{2}\partial-(\Theta^{rq}_{11}-2\Theta^{rq}_{32}+2\Theta^{rq}_{23}+\Theta^{qr}_{44})\nonumber\\
M_{67}&=&(-\Theta^{rr}_{12}-\Theta^{rr}_{34}+\Theta^{rr}_{21}-\Theta^{rr}_{43})\nonumber\\
M_{68}&=&(\Theta^{rq}_{12}+\Theta^{rq}_{34}-\Theta^{rq}_{21}+\Theta^{rq}_{43})\nonumber\\
M_{71}&=&(\Theta^{qr}_{14}+2\Theta^{qr}_{22}-2\Theta^{qr}_{33}+\Theta^{qr}_{41})\nonumber\\
M_{72}&=&(-\Theta^{qq}_{14}-2\Theta^{qq}_{22}+2\Theta^{qq}_{33}-\Theta^{qq}_{41})\nonumber\\
M_{73}&=&2(\Theta^{qr}_{12}+\Theta^{qr}_{21}+\Theta^{qr}_{34}+\Theta^{qr}_{43})\nonumber\\
M_{74}&=&-2(\Theta^{qq}_{12}+\Theta^{qq}_{21}+\Theta^{qq}_{34}+\Theta^{qq}_{43})\nonumber\\
M_{75}&=&-2(\Theta^{qr}_{13}-\Theta^{qr}_{24}+\Theta^{qr}_{31}-\Theta^{qr}_{42})\nonumber\\
M_{76}&=&2(\Theta^{qq}_{13}-\Theta^{qq}_{24}+\Theta^{qq}_{31}-\Theta^{qq}_{42})\nonumber\\
M_{77}&=&-\frac{1}{2}\partial+(\Theta^{qr}_{11}+2\Theta^{qr}_{23}-2\Theta^{qr}_{32}-\Theta^{qr}_{44})\nonumber\\
M_{78}&=&-(\Theta^{qq}_{11}+2\Theta^{qq}_{23}+2\Theta^{qq}_{32}-\Theta^{qq}_{44})\nonumber\\
M_{81}&=&(\Theta^{rr}_{14}+2\Theta^{rr}_{22}-2\Theta^{rr}_{33}-\Theta^{rr}_{41})\nonumber\\
M_{82}&=&-(\Theta^{rq}_{14}+2\Theta^{rq}_{22}-2\Theta^{rq}_{33}+\Theta^{rq}_{41})\nonumber\\
M_{83}&=&2(\Theta^{rr}_{12}+\Theta^{rr}_{21}+\Theta^{qr}_{34}+\Theta^{qr}_{43})\nonumber\\
M_{84}&=&-2(\Theta^{rq}_{12}+\Theta^{rq}_{21}+\Theta^{rq}_{34}+\Theta^{rq}_{43})\nonumber\\
M_{85}&=&-2(\Theta^{rr}_{13}-\Theta^{rr}_{24}+\Theta^{rr}_{31}-\Theta^{rr}_{42})\nonumber\\
M_{86}&=&2(\Theta^{rq}_{13}-\Theta^{rq}_{24}+\Theta^{rq}_{31}-\Theta^{rq}_{42})\nonumber\\
M_{87}&=&(\Theta^{rr}_{11}-\Theta^{rr}_{44}+2\Theta^{rr}_{23}+2\Theta^{rr}_{32})\nonumber\\
M_{88}&=&\frac{1}{2}\partial-(\Theta^{rq}_{11}+2\Theta^{rq}_{23}+\Theta^{rq}_{32}-\Theta^{rq}_{44})\nonumber
\end{eqnarray}
II. Elements of matrix $N$ in equation(88)
\begin{eqnarray}
N_{11}&=&-\frac{1}{2}\partial+(\Theta^{qr}_{11}+\Theta^{qr}_{23}+\Theta^{qr}_{32})\nonumber\\
N_{12}&=&-(\Theta^{qq}_{11}+\Theta^{qq}_{23}+\Theta^{qq}_{32})\nonumber\\
N_{13}&=&(\Theta^{qr}_{13}+\Theta^{qr}_{22}+\Theta^{qr}_{31})\nonumber\\
N_{14}&=&-(\Theta^{qq}_{13}+\Theta^{qq}_{22}+\Theta^{qq}_{31})\nonumber\\
N_{15}&=&(\Theta^{qr}_{12}+\Theta^{qr}_{21}+\Theta^{qr}_{33})\nonumber\\
N_{16}&=&-(\Theta^{qq}_{12}+\Theta^{qq}_{21}+\Theta^{qq}_{33})\nonumber\\
N_{21}&=&(\Theta^{rr}_{11}+\Theta^{rr}_{23}+\Theta^{rr}_{32})\nonumber\\
N_{22}&=&\frac{1}{2}\partial-(\Theta^{rq}_{11}+\Theta^{qr}_{23}+\Theta^{qr}_{32})\nonumber\\
N_{23}&=&(\Theta^{rr}_{13}+\Theta^{rr}_{22}+\Theta^{rr}_{31})\nonumber\\
N_{24}&=&-(\Theta^{rq}_{13}+\Theta^{rq}_{22}+\Theta^{rq}_{31})\nonumber\\
N_{25}&=&(\Theta^{rr}_{12}+\Theta^{rr}_{21}+\Theta^{rr}_{33})\nonumber\\
N_{26}&=&-(\Theta^{rq}_{12}+\Theta^{rq}_{21}+\Theta^{rq}_{33})\nonumber\\
N_{31}&=&(\Theta^{qr}_{12}+\Theta^{qr}_{21}+\Theta^{qr}_{33})\nonumber\\
N_{32}&=&-(\Theta^{qq}_{12}+\Theta^{qq}_{21}+\Theta^{qq}_{33})\nonumber\\
N_{33}&=&-\frac{1}{2}\partial+(\Theta^{qr}_{11}+\Theta^{qr}_{23}+\Theta^{qr}_{32})\nonumber\\
N_{34}&=&-(\Theta^{qq}_{11}+\Theta^{qq}_{23}+\Theta^{qq}_{32})\nonumber\\
N_{35}&=&(\Theta^{qr}_{13}+\Theta^{qr}_{22}+\Theta^{qr}_{31})\nonumber\\
N_{36}&=&-(\Theta^{qq}_{13}+\Theta^{qq}_{22}+\Theta^{qq}_{31})\nonumber\\
N_{41}&=&(\Theta^{rr}_{12}+\Theta^{rr}_{21}+\Theta^{rr}_{33})\nonumber\\
N_{42}&=&-(\Theta^{rq}_{12}+\Theta^{rq}_{21}+\Theta^{rq}_{33})\nonumber\\
N_{43}&=&(\Theta^{rr}_{11}+\Theta^{rr}_{23}+\Theta^{rr}_{32})\nonumber\\
N_{44}&=&\frac{1}{2}\partial-(\Theta^{rq}_{11}+\Theta^{rq}_{23}+\Theta^{rq}_{32})\nonumber\\
N_{45}&=&(\Theta^{rr}_{13}+\Theta^{rr}_{22}+\Theta^{rr}_{31})\nonumber\\
N_{46}&=&-(\Theta^{rq}_{13}+\Theta^{rq}_{22}+\Theta^{rq}_{31})\nonumber\\
N_{51}&=&(\Theta^{qr}_{13}+\Theta^{qr}_{22}+\Theta^{qr}_{31})\nonumber\\
N_{52}&=&-(\Theta^{qq}_{13}+\Theta^{qq}_{22}+\Theta^{qq}_{31})\nonumber\\
N_{53}&=&(\Theta^{qr}_{12}+\Theta^{qr}_{21}+\Theta^{qr}_{33})\nonumber\\
N_{54}&=&-(\Theta^{qq}_{12}+\Theta^{qq}_{21}+\Theta^{qq}_{33})\nonumber\\
N_{55}&=&-\frac{1}{2}\partial+(\Theta^{qr}_{11}+\Theta^{qr}_{23}+\Theta^{qr}_{32})\nonumber\\
N_{56}&=&-(\Theta^{qq}_{11}+\Theta^{qq}_{23}+\Theta^{qq}_{32})\nonumber\\
N_{61}&=&(\Theta^{rr}_{13}+\Theta^{rr}_{22}+\Theta^{rr}_{31})\nonumber\\
N_{62}&=&-(\Theta^{rq}_{13}+\Theta^{rq}_{22}+\Theta^{rq}_{31})\nonumber\\
N_{63}&=&(\Theta^{rr}_{12}+\Theta^{rr}_{21}+\Theta^{rr}_{33})\nonumber\\
N_{64}&=&-(\Theta^{rq}_{12}+\Theta^{rq}_{21}+\Theta^{rq}_{33})\nonumber\\
N_{65}&=&(\Theta^{qr}_{12}+\Theta^{qr}_{21}+\Theta^{qr}_{33})\nonumber\\
N_{66}&=&\frac{1}{2}\partial-(\Theta^{rq}_{11}+\Theta^{rq}_{23}+\Theta^{rq}_{32})
\end{eqnarray}

\section{References:}
\par[1] Bilinear integrable systems:From Classical to Quantum, Continuous to Discrete. L. Fadeev, P. Van Moerbeke, F. Lambert. Springer(2006).\\

\par [2] Integrable Models: A. Das. World Scientific(1989).\\

\par [3] Integrable Systems: S. P. Novikov Cham. Univ. Press(2008).\\

\par [4] A. P. Fordy:J. Phys. A: Math Gen. \textbf{17}(1984)1235.\\

\par [5] Han-Yu Wei and Tie Cheng Xia: J. Math. Phys.\textbf{55}(2014)083501.\\

\par [6] Hai-Yong Ding, Xiang Tian, Xi-Xiang Xu, Hong-Xiang Yang: Mod. Phys. Lett. B \textbf{21} nos 2 and 3(2007)155.\\

\par [7] B. Fuchssteiner: Prog. Theor. Phys. \textbf{65}(1981)861 and Prog. Theor. Phys. \textbf{70}(1983)1508.\\

\par [8] B. Fuchssteiner and A. S. Fokas: Physica D  \textbf{4}(1981)47.\\

\par [9] Y. Zhang: Phys. Lett. A.\textbf{317}(2003)280.\\

\par [10] Bacicai and Gu Zhuquan: J. Phys. A \textbf{24}(1991)963.\\

\par [11] Zhang and H. Tau: Chaos, Solitons, Fractals \textbf{39}(2009)1109.\\

\par [12] Lie Algebraic Method in Integrable systems: A Roy Chowdhury. Chapman and Hall, CRC press.\\

\par [13] Fa Jun: Chin. Phys. B \textbf{21}(2012)010201.\\

\par [14] F. B. Wahlquist and H. D. Estabrook: J. Math. Phys.\textbf{17}(1976)1293. \\

\par [15] A. RoyChowdhury and S. Ahamed: Phys. Rev. D\textbf{10}(1985)2780.\\

\par[16] Problems in Theoretical Physics Vol-2: Willi-Hans Steeb. Wissenschaffsverlag, 1990.\\

\par [17]W. X. Ma and R. G. Zhou: J. Math. Phys. \textbf{40}(1999)4419.\\

\par[18] G. Tu: J. Math. Phys. \textbf{30}(1989)330.

\end{document}